\documentclass[preprint,3p,times,twocolumn,sort&compress]{elsarticle}

\usepackage{amssymb}
\usepackage{amsmath,epsfig}
\usepackage{booktabs} 
\usepackage{tabularx} 
\usepackage[colorlinks=true,linkcolor=black,urlcolor=blue,citecolor=black]{hyperref} 
\usepackage{lettrine} 
\usepackage{graphicx} 
\usepackage{subcaption} 
\usepackage[numbers]{natbib} 
\usepackage{amsmath} 
\usepackage{amssymb} 
\usepackage{lipsum} 
\usepackage{xcolor} 
\usepackage{multirow} 
\usepackage{array} 
\usepackage{tabularx}
\usepackage{booktabs}
\usepackage{arydshln}  
\usepackage{titlesec} 
\titleformat{\paragraph}
{\normalfont\normalsize\itshape}{\theparagraph}{1em}{}
\titlespacing*{\paragraph}
{0pt}{3.25ex plus 1ex minus .2ex}{1.5ex plus .2ex}

\begin{document}

\begin{frontmatter}



\title{Onboard Processing of Hyperspectral Imagery: Deep Learning Advancements, Methodologies, Challenges, and Emerging Trends}



\author[inst1]{Nafiseh Ghasemi $^{*}$}
\author[inst1,inst3]{Jon Alvarez Justo~ }
\author[inst2]{Marco Celesti $^{**}$}
\author[inst1]{Laurent Despoisse~ }
\author[inst1]{Jens Nieke~ }

\affiliation[inst1]{organization={European Space Agency, European Space Research and Technology Centre (ESA-ESTEC)},
                country={The Netherlands.},
            }

\affiliation[inst2]{organization={HE Space for ESA - European Space Agency, European Space Research and Technology Centre (ESA-ESTEC)},
                country={The Netherlands.},
            }
            
\affiliation[inst3]{organization={Department of Engineering Cybernetics, Norwegian University of Science and Technology},
                city={},
                country={Norway. \\ $^{*}$ nafiseh.ghasemi@esa.com, $^{**}$  marco.celesti@ext.esa.int},
            }


\begin{abstract}
Recent advancements in deep learning techniques have spurred considerable interest in their application to hyperspectral imagery processing. This paper provides a comprehensive review of the latest developments in this field, focusing on methodologies, challenges, and emerging trends. Deep learning architectures such as Convolutional Neural Networks (CNNs), Autoencoders, Deep Belief Networks (DBNs), Generative Adversarial Networks (GANs), and Recurrent Neural Networks (RNNs) are examined for their suitability in processing hyperspectral data. Key challenges, including limited training data and computational constraints, are identified, along with strategies such as data augmentation and noise reduction using GANs. The paper discusses the efficacy of different network architectures, highlighting the advantages of lightweight CNN models and 1D CNNs for onboard processing. Moreover, the potential of hardware accelerators, particularly Field Programmable Gate Arrays (FPGAs), for enhancing processing efficiency is explored. The review concludes with insights into ongoing research trends, including the integration of deep learning techniques into Earth observation missions such as the CHIME mission, and emphasizes the need for further exploration and refinement of deep learning methodologies to address the evolving demands of hyperspectral image processing.
\end{abstract}



\end{frontmatter}


\section{Introduction}
\begin{figure*}[htbp]
    \centering
    \includegraphics[width=\textwidth] {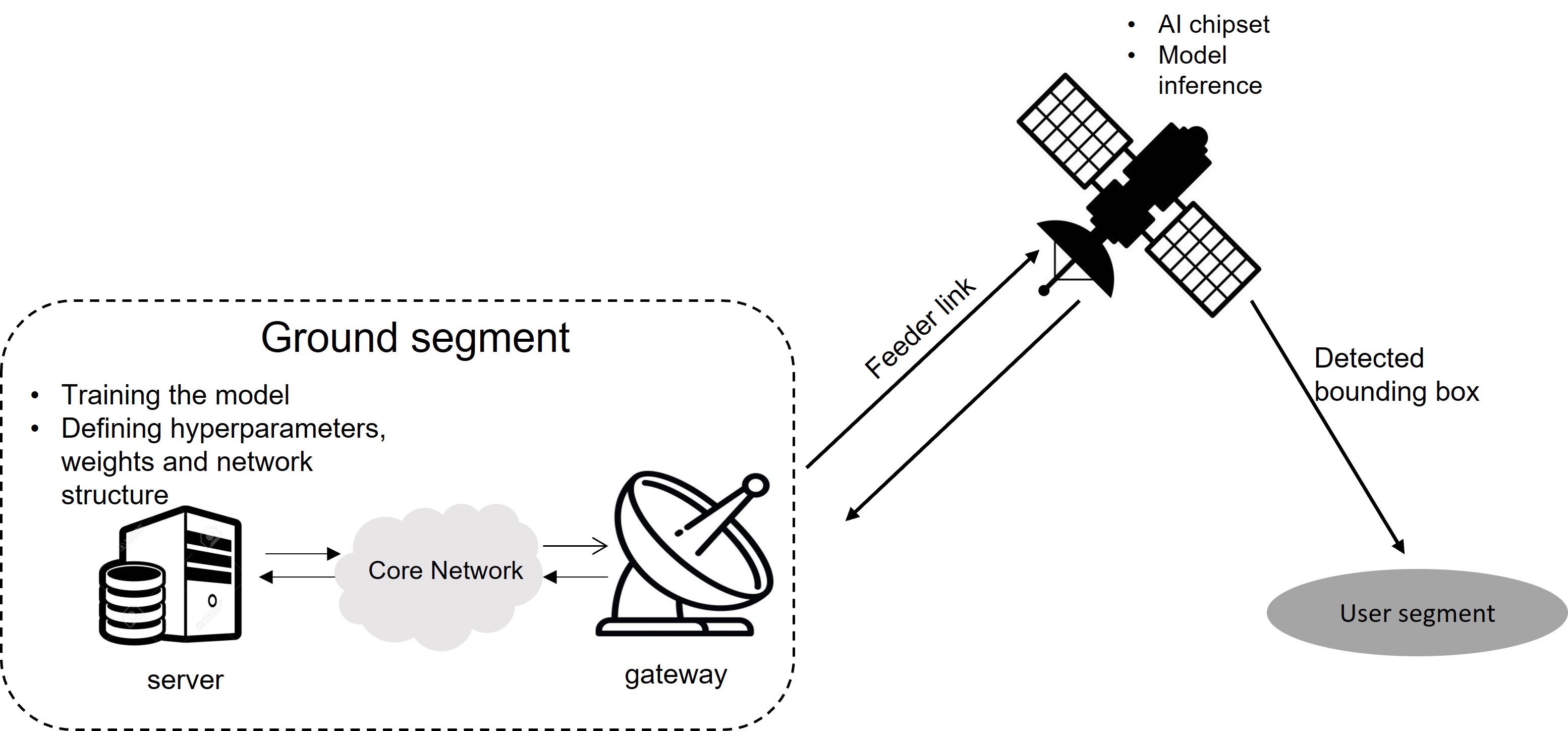}
    \caption{On-board data processing diagram. The AI model will be trained on ground and network structure and hyper-parameters are transferred to the satellite.}
    \label{fig:onboard}
\end{figure*}

Imaging spectroscopy within the visible to short-wave infrared (VSWIR) range of the electromagnetic spectrum is a powerful Earth observation tool that evolved significantly in the last forty years \cite{Rast2019}. In the context of remote sensing, a broad scope of applications and research fields can benefit from the unique capacity of imaging spectroscopy which accurately measures the distinct spectral signatures characterizing the variety of targets on the Earth's surface. To this extent, spectroscopy significantly enhances various monitoring applications in e.g., agriculture, chemical leak detection \cite{zhou2022detection}, security \cite{yuen2010introduction} and oil spills. Furthermore, spectroscopic observations allow advancements in geological classification \cite{peyghambari2021hyperspectral, gao2021generalized}, ocean color studies \cite{grotte2021ocean}, while assisting the assessment of natural hazards before and after their occurrence \cite{lassalle2021monitoring}. Nowadays, hyperspectral (HS) sensors have the capability to capture hundreds of spectral bands spanning across a wide spectrum and with high spectral resolution, hence resulting in highly-dimensional data cubes. Each pixel contained within the data cubes consists of a spectral signature, similar to a unique fingerprint \cite{Shippert2003}, revealing how light interacts with the imaged target of interest. This interaction is presented through variations in reflections and absorption patterns across the different wavelengths. The patterns within each image pixel unveils valuable feature information of the Earth's surface and its quantitative properties, which may be analysed with techniques such as image classification or quantitative retrieval of geophysical properties through e.g., regression techniques. The significance of imaging spectroscopy is emphasized by the numerous satellite missions integrating, for instance, HS sensors in their payloads. A variety of HS missions are nowadays producing a nearly uninterrupted stream of data. As an example, HS technology is incorporated not only in airborne systems such as NASA/AVIRIS-NG \cite{vane1993airborne}, but also in spaceborne missions such as DLR/EnMAP \cite{Guanter2015}, ASI/PRISMA \cite{lopinto2022current}, NASA/EMIT \cite{green2022nasa}, ESA/CHIME \cite{nieke2023copernicus,celesti2022copernicus} and NASA/SBG \cite{sen2023overview} (both missions planned for launch in 2028). In these missions, most of their on-board HS sensors capture data with approximately spectral and spatial resolutions of, respectively, 10 nm and 30 m. When combined with repeated acquisitions in relatively large swaths (30 km to 150 km), the data produced requires large storage and computational processing power. Therefore, considering the rising demands for this Earth observation data, current state-of-the-art efforts focus on the development of automated data analysis solutions to process, already in orbit, the vast amounts of data being generated. To accomplish this automated feature extraction is of key relevance. After preparing and training the extraction models on the ground segment, the models are uplinked to the satellite for near real-time inference on e.g., commercial off-the-shelf hardware processing units embedded in the spaceborne payloads. The process has been schematically presented in Figure \ref{fig:onboard}. The success of the deployment of deep learning algorithms depends on several aspects. For instance, the quality of the data used for training the models, the real acquired data by the satellite, and the limited on-board memory as well as power supply constraints, among other factors.

It is a current trend to develop algorithms that can process data near real-time, and extract only the mission required information to optimize storage and reduce processing costs \cite{Khan2018}, while saving downlink resources.  

In earlier studies, imaging spectroscopy analysis was performed using machine learning (ML) techniques such as K-nearest neighbor \cite{Ma2010}, support vector machines (SVMs) \cite{van2007classifying}, and Gaussian unmixing models \cite{Altmann2013}. Moreover, sparse signal representation methods have been used to classify noisy data with the help of a learned dictionary \cite{Sun2016}. However, in many cases traditional ML techniques that require manual feature extraction are not suitable alternatives, and hence the emerging deep learning (DL) models have found their place within the HS community \cite{Zhang2016, Gao2018}, due to their promising potential to automate the learning of complex non-linear data patterns. DL techniques can detect features that are often not possible to be inferred by humans analyses \cite{He2018}, while allowing solving diverse tasks for on-board data processing from data compression \cite{Sun2016, Lorenzo2020} to target detection \cite{Nalepa2020} and feature extraction \cite{Wang2020}. For instance, DL techniques have made significant advancements in data compression as demonstrated in \cite{gedalin2019deepcubenet}, where for the first time the literature outperforms traditional iterative reconstruction methods in the field of Compressive Sensing (CS) for HS imagery \cite{justo2022comparative}. Additionally, DL techniques are prevalent in other data processing tasks such as pixel-wise classification, which is frequently referred as image classification in the HS field. Conversely, in computer vision, the term commonly used for pixel-level classification is image segmentation. In the case of HS imagery, both terms can be used interchangeably \cite{Nalepa2020, Tarabalka2010}. Pixel-wise classification for HS imagery results into non-overlapping segments also called super-pixels \cite{fang2015spectral} or homogeneous image regions. 

The objective of this paper is to evaluate various DL techniques for on-board processing in terms of reliability, ability to handle noisy data, and network architecture. These factors, among others, play a crucial role in the implementation of DL for on-board applications. This is of relevance for upcoming satellite missions such as CHIME and SBG. Additionally, the study assesses the capability of the networks to be trained with limited training samples. The outcome of this analysis will inform the decision on which network architectures and configurations may be optimal for on-board imaging spectroscopy segmentation.

\section{Deep Learning for Segmentation in Imaging Spectroscopy}

We present first Convolutional Neural Networks (CNNs) with  different data processing approaches from spectral or spatial to a combination of spectral and spatial. Furthermore, other significant architectures we consider are Autoencoders, Deep Belief Networks (DBNs), Generative Adversarial Networks (GANs), and Recurrent Neural Networks (RNNs). These architectures are flexible and adaptable for on-board processing of imaging spectroscopy data. Finally, we discuss about the architectures' challenges as well as new trends to tackle them. 

\subsection{Spectral and Spatial Dimensions in Imaging Spectroscopy Processing}

\begin{figure*}[htbp]
    \centering
    \includegraphics[width=\textwidth] {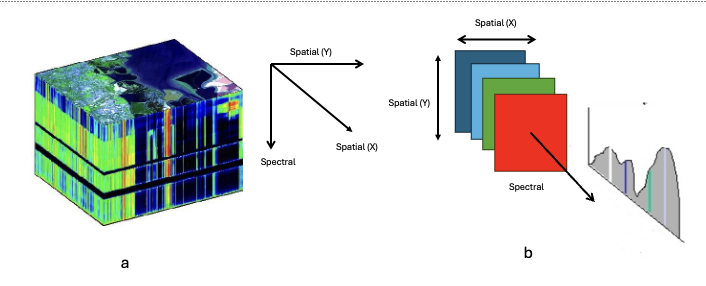}
    \caption{A visualization of spatial and spectral dimensions on a) hyperspectral data cube and b) a schematic representation of hyperspectral pixel.}
    \label{fig:spectrum}
\end{figure*}

HS data can be processed based on any of its different dimensions, either spectral and/or spatial. A diagram showing spectral and spatial dimensions on the hyperspectral data cube is shown in Figure \ref{fig:spectrum}. For instance, in earlier studies for DL methods, a prevalent approach involved pixel-wise processing to extract only the spectral signature from each individual pixel and subsequently comparing it to a known spectral signature of a target object, and hence prior knowledge about the reference target was required (an example is given in \cite{yang2009nonparametric}). However, it is a common practice to process multiple neighboring spatial pixels, rather than just one pixel, with the goal of incorporating additional spatial information to infer each pixel classification. In these scenarios involving the processing of both spectral and spatial domains, it is often more efficient to concentrate on data sub-volumes, represented as 3D patches, rather than handling entire image data cubes, as this can be often computationally more expensive. In previous research, findings from the HYPSO-1 mission \cite{justo2023sea} suggest that in certain scenarios, the segmentation of individual spectral signatures without accounting for the neighboring spatial context can yield superior performance compared to the joint processing of both spectral and spatial domains. However, in any case, in imaging spectroscopy, it is common to reduce the correlated information in the spectral domain and hence remove redundant spectral channels. To achieve this, dimensionality reduction methods such as PCA \cite{fernandez2019fpga}, Kernel-ICA \cite{fong2007dimension, jayaprakash2018dimensionality}, and autoencoders \cite{zabalza2016novel} can be employed. 

\subsection{Convolutional Neural Networks}
   
Artificial Neural Networks (ANNs) are inspired by biological neural systems and consist of: an input layer, one or more hidden layers and an output layer \cite{Cybenko1989}.  When neural networks comprise more than two hidden layers, we commonly refer to them as deep neural networks (DNNs) \cite{Montavon2018}. A subtype of DNNs is CNNs. These convolutional networks have been introduced for the purpose of extracting information from images in computer vision \cite{Albawi2017}. This type of network has found extensive applications in different image analyses tasks \cite{Tajbakhsh2016}. In CNNs, the input image is structured according to the network's architecture, typically arranged in three dimensions: width $(w)$, height $(h)$, and depth $(d)$. In the context of HS imagery, the input depth $(d)$ corresponds to the number of bands per image pixel. On the other side, the input width $(w)$ and height $(h)$ correspond to the spatial dimensions. 
Within each layer in the network, neurons are connected to a selected subset of neurons from the previous layer. This arrangement serves to decrease the number of weights that must be trained \cite{Tajbakhsh2016}.

CNNs have also been combined with ML methods such as SVMs to extract features and increase robustness against overfitting, as demonstrated in \cite{leng2016cube}. In this study, the approach involves assembling a spectral-spatial multi-feature cube by combining the data from a target pixel and its nehigboring spectral information, all without the need for additional modifications to the network's architecture. This combined approach has been employed for land cover classification. 
In another study presented in \cite{yang2016hyperspectral}, a 2-channel deep CNN was employed for land cover classification by combining spectral-spatial features. Additionally, a hierarchical framework was adopted for this task in \cite{wei2017spectral}. Similarly, in \cite{chen2017deep}, a method is introduced to extract spatial and spectral features from imaging spectroscopy and lidar data using a CNN. Furthermore, a pixel-wise classification approach was applied in \cite{xu2017multisource}, using a 2-channel CNN and multi-source feature extraction. To continue, the research in \cite{jiao2017deep} introduces a framework for the feature classification of imaging spectroscopy data. Concretely, it employs a Fully Convolutional Network (FCN) to predict spatial features by analyzing multiscale local information. These spatial features are then combined with spectral features using a weighted method. Subsequently, this approach utilizes SVMs for the classification task.

\subsubsection{Spectral Dimensional CNN (1D-CNN)}

Although 2D CNNs are a common imaging analysis method employed in the field of computer vision, 1D CNNs can also be used for pixel-level classification in imaging spectroscopy. These networks often apply the 1D convolution over the spectral domain. However, they are susceptible to noise, thus posing challenges to use them in remote sensing \cite{hu2015deep}. A possible solution suggested in the literature is to utilize averaged spectrum from a group of neighboring pixels, a method particularly suited for small-scale analyses such as crop segmentation \cite{sun2021supervised}. Another approach is to perform PCA before applying CNNs, offering another strategy to mitigate noise-related issues. A method used to reduce dimensionality of HS imagery is Minimum Noise Fraction (MNF) \cite{green1988transformation}. The MNF transform involves two sequential operations as presented next. First, it estimates the noise in the data using a correlation matrix and subsequently scales it based on variance. This operation decorrelates and reduces noise within the data without considering inter-band noise. Next, the second operation in the MNF transformation takes into account the original correlations and creates components that incorporate weighted information for the variance across all bands within the original dataset. MNF has found application in tasks such as denoising and image classification \cite{wang2006independent,nielsen2010kernel,sharma2022aviris}. Typically, the MNF algorithm requires more computational resources compared to other commonly employed HS processing techniques such as PCA or Linear Discriminant Analysis (LDA) \cite{bandos2009classification}. Therefore, MNF may not be prioritized for on-board image processing  tasks. An alternative solution, as detailed in \cite{he2017multi}, involves the utilization of a multi-scale CNN applied to a pyramid of data encompassing spatial features at multiple scales. In case of limited training observations, a different strategy proposed in \cite{li2017hyperspectral} consists of performing band selection  before CNN inference.

\subsubsection{Spectral-Spatial Dimensional CNN (2D-CNN)}
In many cases, utilizing both spectral and spatial features typically leads to better results in imaging spectroscopy processing. In \cite{zhang2017spectral}, a dual-stream channel CNN was employed. This network extracts spectral features using the approach in \cite{hu2015deep}, while spatial features are derived from the method in \cite{slavkovikj2015hyperspectral}. These extracted features are subsequently integrated using a softmax classifier.

In \cite{chen2016deep}, a combination of the $l_2$ norm and sparse constraints, incorporating a similar combination of spectral-spatial features, is employed. In other studies, AlexNet \cite{iandola2016squeezenet} has been employed for spatial-spectral analyses, involving architectures like Densenet and VGG-16 \cite{jiao2017deep, zhi2019dense}. Furthermore, other works \cite{liu2018deep} adopted a few-shot learning approach proposed in \cite{li2006one}. This approach has been used to learn a metric space, promoting the proximity of samples belonging to the same class. This was done in order to address the challenge posed by a limited number of training smaples.

An alternative strategy to improve accuracy while having a shortage of training data has been proposed in \cite{paoletti2018deep}. In this approach, the redundant information within the hidden layers is explored to identify connections and improve the training process. Additional examples of combining spatial-spectral features to improve the learning process can be found in various works \cite{makantasis2015deep,yue2015spectral, achanta2012slic}. These studies have used a variety of methods based on super-pixel reconstruction of different features, aiming to improve the accuracy of segmentation. In addition to these methods, other works \cite{mei2017learning} have introduced a technique known as sensor-based feature learning consisting of the reconstruction of five layers of spectral-spatial features tailored to the specifications of the sensor used, contributing to the improvement of the learning process. An additional improvement to sensor-based training was given in \cite{fang2018hyperspectral}, which uses a novel architecture that actively transforms input features into meaningful response maps for classification tasks. All the aforementioned studies have used complex multi-step procedures, rendering them not suitable for near-real-time processing in imaging spectroscopy. However, these studies demonstrate that the adoption of multi-scale and multi-feature approaches has been substantiated as a means of achieving improved performance, as emphasized in the mentioned studies.


\begin{figure*}[htbp]
    \centering
    \includegraphics[width=\textwidth] {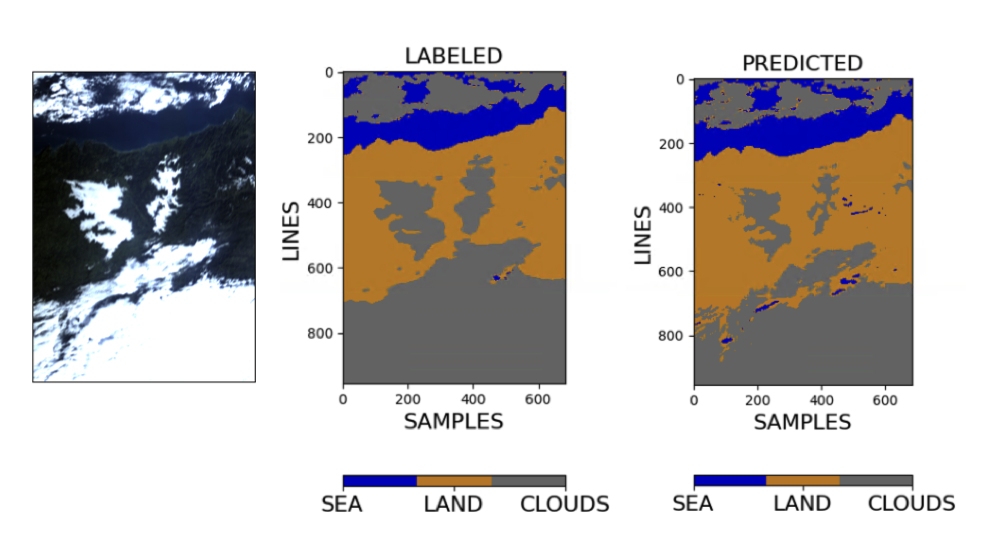}
    \caption{1D-CNN results on a sample dataset from HYPSO-1 cubesat captured over Spain \cite{justo2023sea}.}
    \label{fig:HYPSO-1}
\end{figure*}

There have been studies to compare both 1D vs 2D CNNs for on-board deployment. In one study 1D and 2D have been test on HYPSO-1 cube sat launched by Norwegian University of Science and Technology (NTNU) \cite{justo2023sea}. They evaluate  classification performance and model parameters. An example of the results is shown in Figure \ref{fig:HYPSO-1}. The classification task involves identifying oceanic, terrestrial, and cloud formations. The results have shown advantage of using 1D-CNN for on-board processing. The 1D-CNN architectures underwent additional testing in \cite{kovac2024deep} using data from NASA's EO-1 Hyperion and Ziyuan-1 satellites, confirming the promising results of these architectures.

\begin{figure*}[htbp]
    \centering
    \includegraphics[width=\textwidth] {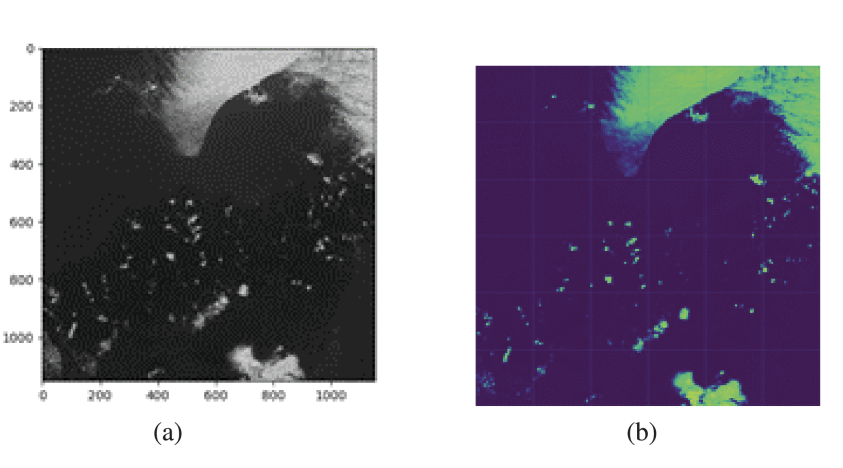}
    \caption{Captured image over Etna Volcano, Italy by Phi-Sat-1, a)original image, b)neural network detection results from on-board processor unit \cite{giuffrida2021varphi}.} 
    \label{fig:phi-sat-1}
\end{figure*}

Another example is the Phi-Sat-1 mission, an initiative under the European Space Agency (ESA) represents a pioneering effort in deploying artificial intelligence (AI) directly onboard, leveraging a dedicated chip for deep convolutional neural network (CNN) inference \cite{giuffrida2021varphi}. 1D and 2D-CNN have been tested on-board for cloud detection. It demonstrates the feasibility of onboard CNN for cloud detection, achieving high accuracy rates more than 90\% by 1D CNN. An example of the results of Phi-Sat-1 is shown in Figure \ref{fig:phi-sat-1}.

It shows 
 Following the success of the Phi-Sat-1 mission. Phi-Sat 2 mission has been proposed aiming to integrate cutting-edge AI technology into Earth Observation  missions \cite{marin2021phi}. Phi-Sat-2 will demonstrate the deployment of AI applications onboard, enabling functionalities. This initiative represents a pivotal step in advancing innovative techniques to meet user-driven science and application needs.

\subsection{Autoencoders}
To deal with the issue of limited training data for processing imaging spectroscopy, various approaches based on autoencoders with different variations have been tested. For the first time \cite{lin2013spectral}, PCA was applied in the spectral dimension while auto-encoders were applied at the same time to the other two spatial dimensions, resulting in improved feature extraction for classification. Subsequent research in \cite{chen2014deep} and \cite{tao2015unsupervised} utilized stacked autoencoders in combination with PCA to flatten the spectral dimension, followed by the utilization of SVM and Multi-Layer Perceptron (MLP) to perform classification. Furthermore, \cite{zhang2019stacked}  focused on optimizing stacked autoencoders specifically for anomaly detection within imaging spectroscopy. A combination of auto-encoders and CNNs have also been tested in multi-scale approaches for feature extraction  \cite{yue2016deep}. Another important advantage of using stacked autoencoders is their effectiveness in handling noisy input data. As illustrated in \cite{liu2015hyperspectral}, a stacked autoencoder was utilized to generate feature maps from noisy input data, followed by the application of super-pixel segmentation and majority voting for further processing. In a different study, \cite{xing2016stacked} used a pre-trained network consisting of stacked encoders combined with logistic regression to perform supervised classification on noisy input data. Furthermore, a framework based on stacked autoencoders was initially proposed in \cite{windrim2017physics} to perform also unsupervised classification on noisy input data. This framework was later improved as an end-to-end classification pipeline tailored for imaging spectroscopy in \cite{ball2018deep}.

\subsection{Deep Belief Networks, Generative Adversarial Networks, and Recurrent Neural Networks}
Deep Belief Networks (DBNs) have the capability of reducing dimensions, and hence they are a suited alternative for feature extraction tasks. In \cite{chen2015spectral}, a DBN was employed combined with logistic regression to perform feature extraction. Furthermore, a combination of one- and two-layer DBN was combined with PCA for feature extraction. To perform near real-time anomaly detection, DBNs have been investigated and tested and have shown to deliver promising results in the extraction of local objects, as demonstrated in \cite{ma2018unsupervised}. Another approach proposed by \cite{huang2019hyperspectral} consisted of combining DBNs with wavelet transforms. Additionally, other works \cite{wang2019nonlinear, ozkan2018endnet} utilized DBNs for unsupervised classification. In the later study, an end-to-end classification framework based on DBNs and the spectral angle distance metric was proposed.

As regards Generative Adversarial Networks (GANs), these networks employ a pair of competing neural networks, one as a generator and  the other as a discriminator \cite{creswell2018generative}. These networks have found usefulness to perform classification when dealing with limited training data \cite{he2017generative}. In similar scenarios, GANs have been employed to perform the classification of imaging spectroscopy utilizing the discriminator network for this purpose \cite{zhang2018unsupervised, zhan2018semi, bashmal2018siamese}.

Finally, Recurrent Neural Networks (RNNs) are mainly used for the analysis of time series data. However, in specific applications, they can be adapted to process video series as sequences of data (with each spectral band representing a sequence) , enabling RNNs to identify temporal similarities within these frames \cite{wu2017convolutional,mou2017deep}. A novel approach was proposed in \cite{liu2017bidirectional}, where a combination of RNNs was utilized to explore the spectral domain, while Long Short-Term Memory (LSTM) networks were used to explore spatial features. Furthermore, RNNs have been used to process mixed pixels in the spectral dimension, particularly in situations affected by noise \cite{shi2018superpixel}.

\subsection{Unsupervised and Semi-Supervised Approaches}
Due to the common challenge of dealing with a limited pool of available training samples, semi-supervised and unsupervised approaches are getting more popular in the field of imaging spectroscopy. Illustrative examples can be found in studies such as \cite{ratle2010semisupervised} and \cite{romero2015unsupervised}, where semi-supervised techniques and layer-wise classification have been adopted to process extensive sets of imaging spectroscopy data. An additional example of pixel-wise classification can be found in the research conducted in \cite{maggiori2016convolutional}, which used an unsupervised method with CNNs. Initially, the model was trained using imprecise training samples, and subsequently, classification accuracy was improved by incorporating a small set of accurately labeled training samples. In \cite{mou2017unsupervised}, the challenge of limited training samples was addressed by employing a convolution-deconvolution network for unsupervised spectral-spatial feature learning. Particularly, the convolution network was applied to reduce dimensionality, while the deconvolution stage was utilized to reconstruct input data. Another strategy to address the lack of training samples involves improving the training procedure as explained in \cite{wu2017semi}. In this work, unlabeled data is used in combination with a few labeled samples, and an RNN is used for the classification of imaging spectroscopy data. Finally, another method that was tested involved the utilization of ResNet to learn spectral-spatial features from unlabeled data, showing promising results in \cite{feng2019multisource}.

\subsection{Challenges and Emerging Trends in Imaging Spectroscopy Processing}
\subsubsection{The Training Data Gap in Imaging Spectroscopy}

Advances have been made to improve the volume of training data for imaging spectroscopy. For instance, a recent study in \cite{justo2023open} introduced A labeled dataset, comprising 200 HS images captured from the HYPSO-1 \cite{bakken2023hypso} covering sea, cloud and land to help training onboard segmentation algorithms. Another example is the dataset captured by EMITS. This dataset includes different images from various land covers all around the world \cite{green2023performance}. 
There are other datasets which have been extensively used in previous studies i.e. university of Pavia, indian pines, Huston and Botsowana can be mostly found in \cite{hyperspectraldatasets}.

However, despite these efforts to increase the training data, there is still an important lack of training datasets, coupled with a scarcity of ground-truth labels due to the exceptionally time-consuming nature of the labeling process. Consequently, dealing with a shortage of training data still persists as a challenge also in the field of imaging spectroscopy. 
Novel approaches have been explored to address this challenge, including the utilization of semi-supervised techniques \cite{pan2018mugnet}, self-supervised methods based on solving previous proxy tasks \cite{ghamisi2016self}, and domain adoption techniques \cite{wang2019domain}. Additionally, another approach known as active transfer learning has been studied, using the most discriminative features from unlabeled training observations \cite{liu2016active}. 

\subsubsection{Management of Noisy Data}
Several approaches have gained attention to reconstruct high-quality input data for classification. One important work studies the application of super-resolution techniques combined with transfer learning, with the aim to reduce noise and improve the quality of input training samples \cite{yuan2017hyperspectral}. Other studies have used CNNs with sparse signal reconstruction \cite{lin2018dual}, and have employed the Laplacian Pyramid Network (LPN) \cite{he2018hyperspectral} for enhancing input data with lower noise. Another alternative explored in \cite{xie2019high} uses structure tensors with a deep CNN to improve data quality and diminish the noise level.

\subsection{Improving Speed and Precision}
A new trend within the field of computer vision involves the adoption of CapsuleNets (CapsNet) \cite{roy2018effects}. Capsule Networks  employ a series of nested neural layers, enabling them to improve model scalability while simultaneously accelerating computational speed. Examples of CapsNets can be found in \cite{paoletti2018capsule, wang2018hyperspectral, zhu2019deep}. The utilization of spectra-spatial CapsNet has shown fast convergence capabilities, mitigating effectively overfitting \cite{yin2019hyperspectral}.

\subsection{Hardware Accelerators}
To enhance the efficiency of HS data processing, different hardware platforms have been explored, such as computing clusters \cite{Wu2016}, Graphics Processing Units (GPUs) and Field Programmable Gate Arrays (FPGAs) \cite{Gonzalez2013}. Recent advancements in FPGA technology have positioned them as a suitable option for performing on-board image processing in both airborne and spaceborne platforms \cite{Plaza2011}. An FPGA is a hardware programmable component consisting of an array of logic blocks, Random Access Memories (RAMs), hardcopy IPs, Input/Output (I/O) pads, routing channels, among other building blocks such as specialized Digital Signal Processing (DSP) units \cite{Kuon2008}. The fabric logic can be customized to execute different functions at different times and levels. A previous generation of similar technology was named ASICs from Application-Specific Integrated Circuits \cite{Mittal2008}. However, FPGAs offer greater flexibility while achieving minimal power consumption and high performance compared to other platforms for on-board processing of HS imagery \cite{gonzalez2011fpga}. Some studies have explored on-board HS processing using FPGAs, including tasks such as data compression and image segmentation \cite{Lopez2013}. In preivous versions, FPGAs were employed for tasks such as end-member extraction \cite{gonzalez2011fpga}, as exemplified in \cite{bernabe2011fpga}. In particular, the Xilinx Virtex-5 FPGA was utilized for automatic target detection. Moreover, FPGAs were used to perform end-member extraction for multiple targets \cite{Lei2019}. Furthermore, experiments involving spectral signature unmixing have also been conducted on both FPGAs and GPUs, yielding competitive results in terms of accuracy. In one particular study, an FPGA was utilized to demonstrate its on-board processing capabilities to detect chemical plumes \cite{Theiler2018}. This study served as a pilot phase in the development of  an Artificial Intelligce unit intended for use in an upcoming HS satellite mission to be launched by NASA's Jet Propulsion Laboratory (JPL).

However, a main drawback of design in FPGA platforms is the often difficulty of its configuration and programming. To address this, Intel has developed the OpenCL package, while Xilinx has introduced Vitis to automate the synthesis of low-level Register-Transfer Level (RTL) code needed to program the FPGA fabric logic \cite{Wang2022}. Consequently, there have been limited studies exploring the implementation of DL on FPGAs. As a result,  future studies should  involve the implementation of DNNs on FPGAs within the proposed hardware architecture for future Earth observation missions \cite{Omar2021}.

\section{Discussion and Summary}
We have investigated the most recent developments in the application of deep learning to hyperspectral imagery. Nearly all of the reviewed studies emphasized the scarcity of training data as a primary limiting factor, hindering the adoption of deep learning widely in the HS image processing field. Another important obstacle was the limiting computational infrastructure and hardware, particularly in remote sensing. Our study revealed that while there is a substantial body of studies and work on using deep learning for land cover classification, there remains a critical gap in research pertaining to target and anomaly detection, data fusion and spectral unmixing, and deep learning on FPGA platforms.

Network architectures such as UNet, ResNet, and VNet have proven their efficacy as promising starting points. However, further refinement and specification based on application-specific contexts are still warranted. As regards classification of imaging spectroscopy, deep learning has proven to be effective. Nevertheless, the resource-intensive nature of deep learning, together with the satisfactory outcomes achieved already through conventional classification techniques such as SVMs, has led to hesitance among many users when considering the adoption of deep learning. 

Addressing the challenges posed by limited training datasets and noisy input data, employing using Generative Adversarial Networks (GANs) emerges as a feasible solution to generate augmented datasets and reduce noise within the training samples. Additionally, reinforcement learning stands as a promising alternative that merits further research.

\begin{figure*}[htbp]
    \centering
    \includegraphics[width=\textwidth] {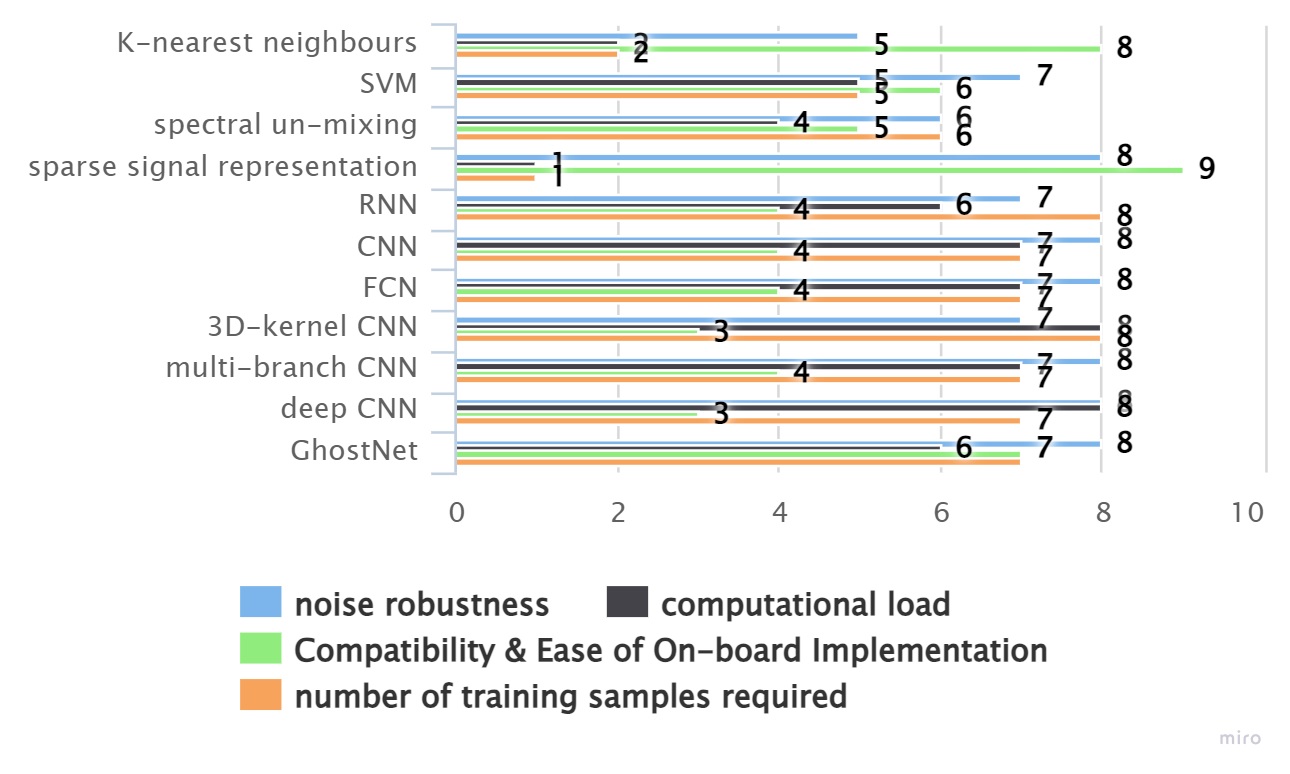}
    \caption{Effectiveness of various on-board analysis methods, with a scale from 0 to 10 representing their suitability in terms of different criteria.}
    
    \label{fig:Chart_V2}
\end{figure*}

Several versions of neural networks have been reviewed, where deep CNN and 3D-kernel-CNN networks have demonstrated highly promising results. Given the focus on optimizing network structures for on-board processing, GhostNets may be a viable alternative, although these networks may not yield the most optimal accuracy. Furthermore, other studies have conducted extensive testing on lightweight CNN models for deployment aboard satellites, comparing 1D and 2D CNNs in which 1D CNNs consistently outperformed 2D CNNs.

When considering the challenges associated with on-board processing in imaging spectroscopy, two key issues arise: noisy data and the scarcity of atmospheric correction at level zero data together with limited training datasets. However, despite the efforts to increase the volume of training data, it still remains insufficient for adequately training models. Consequently, it is crucial to focus on the assessment and testing of different network structures, simulated data, and methods such as self-supervised learning to overcome the lack of training data. To finalise, on-board processing of hyperspectral imagery is the new domain of study that will open many new possibilities in the remote sensing domain and the automation of hyperspectral satellite payloads. In the new CHIME mission there is also ongoing research to perform target detection on board \cite{ghasemi2023feasibility}. Therefore, since there is a trend towards on-board processing of imaging spectroscopy, encompassing both remote sensing and non-remote sensing applications with the aid of hardware accelerators, a comprehensive summary of the prevailing methods, categorized by their suitability for on-board implementation, is provided in Figure ~\ref{fig:Chart_V2}. According to the figure, conventional methods are characterized by ease of implementation, but require a substantial number of training samples and traditional processing and updating procedures. Therefore, within the hyperspectral community, there has been a shift towards employing CNN-based methods.

\section{Conclusion}
In this review, we have explored recent advancements in the application of deep learning techniques to hyperspectral imagery processing. A recurring theme throughout the examined studies has been the challenge posed by limited training data, which continues to hinder widespread adoption of deep learning in hyperspectral image processing. Additionally, constraints in computational infrastructure, particularly in remote sensing applications, have emerged as a significant obstacle.

While deep learning models such as UNet, ResNet, and VNet have demonstrated promising results, further customization and refinement tailored to specific application contexts are warranted. Despite the effectiveness of deep learning in land cover classification, concerns about its resource-intensive nature, coupled with the satisfactory performance of conventional techniques like Support Vector Machines (SVMs), have led to cautious adoption among practitioners.

Addressing challenges related to limited training datasets and noisy input data, the utilization of Generative Adversarial Networks (GANs) for data augmentation and noise reduction presents a viable solution. Additionally, the exploration of reinforcement learning holds promise and merits further investigation.

Various neural network architectures, including deep CNNs and 3D-kernel CNNs, have exhibited promising outcomes. Considering the focus on optimizing network structures for onboard processing, lightweight CNN models and 1D CNNs have shown consistent advantages over their counterparts. However, the issues of noisy data and the scarcity of atmospheric correction at level zero data, along with limited training datasets, persist as primary concerns in onboard processing for hyperspectral imagery.

To overcome these challenges, it is imperative to continue assessing and testing different network structures, simulated data, and methodologies such as self-supervised learning. The trend towards onboard processing of hyperspectral imagery, encompassing both remote sensing and non-remote sensing applications with the support of hardware accelerators, indicates a promising frontier for future research and application. Ongoing efforts in missions like CHIME underscore the growing interest and potential in onboard processing, particularly in tasks such as target detection. In conclusion, this review underscores the evolving landscape of hyperspectral image processing, highlighting the transition towards deep learning-based methods and the importance of addressing key challenges for enhanced operational efficiency and performance.

\section*{Author Contributions}
Nafiseh Ghasemi, is a researcher at European Space Agency (ESA) working on artificial intelligence for on-board processing of hyperspectral data, who is the main author and has done most part of the research project.
Jon Alvarez Justo, is an expert visitor at ESA who was responsible for rewriting, reviewing and adding content about the new hyperspectral missions. 
Marco Celesti, is  a mission scientist at ESA supporting activities for copernicus missions and providing content about hyperspectral applications. 
Laurent Despoisse is spacecraft manager for hyperspectral missions at ESA supporting information on the status of near future CHIME mission.
Jens Nieke, as the project manager of the CHIME group, provided support for the research project and contributed to the overall direction of the study.

\section*{Acknowledgements}
The authors express their gratitude to the European Space Agency (ESA) for providing us support for our research.

\section*{Conflicts of Interest} The authors declare no conflicts of interest.

\vspace{-10pt}
\bibliography{library_1} 



\end{document}